# Lanthanide upconversion nonlinearity: a key probe feature for background-free deep-tissue imaging

Niusha Bagheri[a], Chenyi Wang[b], Du Guo[a], Anbharasi Lakshmanan[a], Qi Zhu[a], Nahid Ghazyani[c], Qiuqiang Zhan[b], Georgios A. Sotiriou[d], Haichun Liu*[a], Jerker Widengren*[a]

[a] *Experimental Biomolecular Physics, Department of Applied Physics, KTH Royal Institute of Technology, SE-106 91, Stockholm, Sweden*

[b] *Centre for Optical and Electromagnetic Research, South China Academy of Advanced Optoelectronics, South China Normal University, Guangzhou 510006, P. R. China*

[c] *Faculty of Physics, Kharazmi University, Tehran, Iran.*

[d] *Department of Microbiology Tumor and Cell Biology Karolinska Institute, SE-171 77, Stockholm, Sweden*

*Corresponding authors: haichun@kth.se; jwideng@kth.se

## Abstract

Lanthanide-based upconversion nanoparticles (UCNPs) have attracted considerable attention in biomedical applications, largely due to their anti-Stokes shifted emission enabling autofluorescence-free signal detection. However, residual excitation light can still interfere with their relatively low brightness. While commonly used lock-in detection can distinguish weak signals from substantial random background, concurrently modulated residual excitation light is not eliminated. This remains a challenge, particularly under demanding experimental conditions.

Here, we explore the inherent nonlinear response of UCNPs and discover that UCNPs can act as frequency mixers in response to intensity-modulated excitation. Particularly, modulated excitation with more than one base modulation frequency can generate additional low-frequency beating-signals. We show how these signals are resolvable by low-speed detectors such as cameras, are devoid of ambient and residual excitation light, and how they can be enhanced through nanoparticle engineering. Detection of beating-signals thus provides a strategy to significantly enhance signal-to-background conditions in UCNP-based bioimaging and biosensing.

## Introduction

Lanthanide-based upconversion nanoparticles (UCNPs) have attracted significant attention for biomedical imaging and biosensing applications.[1,2] The large anti-Stokes shifts in their emission are particularly advantageous, enabling autofluorescence-free detection,[3] potentially offering more favourable signal-to-background conditions and greater sensitivity compared to organic fluorophores with Stokes-shifted emission. However, the relatively low brightness of UCNPs remains a significant obstacle, especially in challenging conditions, such as in small-animal imaging. Although tissue autofluorescence can be effectively eliminated using UCNPs together with appropriate emission filters, residual excitation laser light often remains an issue. Additional filters in the detection system to further reject laser light may not fully resolve this issue because filters are typically designed to achieve their nominal optical density value with normal light incidence,[4] a condition that is not always met in large-area detection scenarios. Therefore, it is essential to develop approaches to enhance the signal-to-background ratio (SBR) in UCNP imaging based on other, orthogonal principles to fully realize their potential in biological applications.

Lock-in detection, based on time-modulated light excitation combined with phase-sensitive detection at the modulation frequency, makes it possible to capture weak signals and dramatically increase detection sensitivity by several orders of magnitude. This technique is widely employed in applications requiring ultrasensitive optical detection, such as in remote optical sensing and in different optical microscopy and spectroscopy techniques.[5-7] Lock-in detection works by distinguishing the periodic signal generated by the modulated light source, effectively discriminating this signal from random background noise. However, lock-in detection does not resolve the issue of residual excitation light, which gets modulated at the same frequency as the optical signal being detected, and is thus not suppressed in the detection.

In this work, we show how excitation modulation and lock-in detection, specifically tailored to take advantage of the inherent nonlinearity of upconversion luminescence (UCL) in response to excitation intensity, can overcome this drawback. By simulations and experiments we first demonstrate how additional frequency components of the UCL signal arise, not present in the residual excitation light background, and how lock-in detection at these frequencies can significantly suppress background light and enhance SBR. We then show how excitation modulation at two different frequencies allows lock-in detection at their difference (beating) frequency, opening for lock-in detection using low time-resolution cameras, and how the relative strength of this beating frequency signal can be optimized by composition and design of the UCNPs. Finally, we show how the proposed lock-in detection approach together with optimized UCNPs enhance SBR in imaging experiments, under high background conditions, and deep into tissues.

**Results and discussion**

We first envisaged our rationale for exploring background-free detection approaches by simulations (Fig. 1), comparing the emission response to sinusoidal and square-wave excitations of UCNPs, with that of an organic fluorophore with Stokes-shifted emission, modelled as depicted in Fig. 1a (equations given in Supporting Information (SI), section S2). When the fluorophore is exposed to a sinusoidal excitation at frequency $f$ it maintains a linear response to the excitation light within the range of excitation intensities applied (maximally 100W/cm$^2$, resulting in an excitation rate ($1.2\times10^7$s$^{-1}$) still much smaller than the decay rates of the excited singlet states and with negligible buildup of dark triplet states[8-12]). The fluorescence signal is then also manifested as a sinusoidal wave at the same frequency as the excitation (Fig. 1b). When the fluorophore is subject to square-wave excitation at frequency $f$ with a 50% duty cycle, the resulting fluorescence similarly follows the excitation pattern (Fig. 1d). As confirmed in the power spectrum of the fast Fourier transform (FFT), no additional frequency components were generated in the fluorophore emission, other than those present in the excitation (Figs. 1c and 1e) (Due to the relatively short (nanoseconds) fluorescence lifetime of most fluorophores, lifetime-related phase lags of the fluorescence relative to the excitation are negligible at the frequencies applied in this study ($< 200$Hz) [13]).

Unlike organic fluorophores, UCNPs exhibit an inherent nonlinear response to excitation intensity, also under non-saturation conditions, fundamentally rooted in their sequential multi-step excitation process preceding their emission.[14,15] We hypothesized that when UCNPs are exposed to a modulated excitation intensity at a certain frequency, they may thus generate emission signal with additional frequency components, beyond those contained in the excitation light. To verify this hypothesis, we conducted similar numerical simulations as for the fluorophore, based on a representative electronic state model of an UCNP, including a sensitizer and an activator yielding a two-photon energy-transfer upconversion system (Fig. 1f) (equations detailed in section S3 in the SI). Upon sinusoidal excitation at frequency $f$, and in contrast to the fluorescence from the organic fluorophore, a second-harmonic signal ($2f$) was observed in the FFT power spectrum of the UCL (Figs. 1g-h), confirming our prediction. Further, for square-wave excitation at frequency $f$ (50% duty cycle) the FFT power spectrum of the UCL also revealed even-order harmonics (Figs. 1i-j), with the $2f$ signal reaching approximately 10% of the strength of the base modulation frequency signal. Comparing the FFT power spectrum of the UCL and that of the excitation suggests that lock-in detection at even-order harmonic frequencies, particularly at $2f$, can offer a background-free signal, also eliminating the influence of modulated residual laser light in imaging experiments.

To experimentally verify the numerical predictions, we first investigated the emission response of NaYF$_4$:20%Yb$^{3+}$, 0.5%Tm$^{3+}$@NaYF$_4$ (**YbTm@Y**) UCNPs, with a well-established upconversion pathway (Fig. 2a),[16] when subject to square-wave excitation (50% duty cycle) using the optical setup illustrated in Fig. S1. Fig. 2b shows a typical excitation intensity time-trace applied in these experiments (modulated at 20Hz), with its FFT power spectrum including odd harmonics (20, 60, 100Hz, …) of the base frequency (20Hz) with negligible even harmonics (Fig. 2c). We then analyzed the generated UCL (430-850nm) (Fig. 2d) and its FFT power spectrum. Notably, significant even harmonics (at 40, 80, 120Hz, …) were observed alongside the original odd harmonics (Fig. 2e), fully validating the predictions from our simulations (Fig. 1).

We then varied the base modulation frequency and observed how it affected the even harmonics, specifically the $2f$ signal. To quantify the relative strength of this signal, we used the second-harmonic weight (SHW), defined as follows:

$$\text{SHW} = \frac{I_{2f}}{(I_f + I_{3f})/2} \times 100\%, \tag{1}$$

where $I_x$ (for $x = f, 2f, 3f$) represents the amplitude at frequency $x$ in the FFT power spectrum. For comparison, the same quantitative analysis was also performed on the excitation light. As shown in Fig. 2f, the SHW of the UCL (430-850nm) increases with $f$, until reaching a maximum, and then decreases with yet higher $f$. For **YbTm@Y** UCNPs, the SHW of the UCL peaks at approximately 16%, at a modulation frequency of 50Hz. Throughout the tested range of $f$, the SHW of the UCL is significantly higher than that of the laser excitation, ruling out the possibility of instrumental artefacts leading to imperfections in the excitation modulation.

Next, we investigated the influence of excitation intensity on the UCL SHW, applying a fixed base modulation frequency at 40Hz and varying the excitation intensity between 4-64W/cm². As shown in Fig. 2g, the UCL SHW decreases monotonically with increasing excitation intensity. These results support the hypothesis that upconversion nonlinearity is the key factor in generating the second harmonic in the UCL, consistent with a lower extent of UCNP nonlinearity at higher excitation intensities due to upconversion saturation.[15]

Experimental investigations on NaYF$_4$:20%Yb$^{3+}$, 2%Er$^{3+}$@NaYF$_4$ (**YbEr@Y**) UCNPs also confirmed the generation of even harmonics in the UCL (Fig. S2a-e). These UCNPs displayed similar dependencies, with some variations in response to the excitation modulation frequency (Fig. S2f) and intensity (Fig. S2g) applied.

These results indicate that even harmonics of the UCL, particularly the $2f$ signal, could offer unique lock-in detection channels for UCNPs, allowing detection of upconversion signals free from background noise and residual excitation light. However, a significant drawback is that these overtones occur at higher frequencies, which according to the Nyquist theorem may not be resolved by low-speed detectors such as cameras. This drawback prompted us to further explore the ability of UCNPs to generate new, lower frequency components.

We realized that the inherent nonlinearity of UCNPs could enable them to function as modulation frequency mixers, generating difference and sum frequency signals when multiple base modulation frequencies are simultaneously applied to the excitation light. This behavior is analogous to that of nonlinear optical crystals in the optical frequency range.

We then studied **YbTm@Y** UCNPs under excitation of two 980-nm lasers modulated at different frequencies ($f_1$ and $f_2$), using a modified optical setup (Fig. S3). Figures 3a and 3c show the recorded excitation and UCL (430-850nm) when the laser excitation beams were separately modulated, at 51 and 53Hz respectively. The FFT power spectrum of the (merged) excitation light reveals the two base modulation frequencies along with their corresponding odd harmonics (the third harmonic shown in Fig. 3b), as expected, with no other detectable frequency components. However, in the FFT power spectrum of the UCL (Fig. 3d) the difference frequency (or the beating frequency - BF, at 2Hz) and the sum frequency (at 104Hz) signals were also clearly detected, along with the $2f$ signals of the two base modulation frequencies (at 102Hz and 106Hz), and in addition to the expected base frequency components.

Next, we investigated the beating and sum frequency signals upon simultaneously varying the two base modulation frequencies while maintaining a fixed BF of 2Hz. To quantify the relative strength of the BF signal, we defined the beating frequency weight (BFW) as follows:

$$\text{BFW} = \frac{I_{f_2 - f_1}}{(I_{f_1} + I_{f_2})/2} \times 100\%, \tag{2}$$

where $I_y$ (for $y = f_1, f_2, f_2 - f_1$) represents the amplitude at frequency $y$ in the FFT power spectrum. Figure 3e shows the dependence of BFW on the average modulation frequency, $(f_1 + f_2)/2$. Within the tested modulation frequency range, the BF signal is consistently generated, with BFW values of a few percent. A similar quantitative analysis of the sum frequency signal reveals a comparable relative strength (Fig. S4).

We then separated the two-photonic 800nm emission band of $Tm^{3+}$ from multi-photonic emissions (430–750nm) and investigated their responses to dual base frequency modulation (21 and 23Hz). It was found that much stronger BFWs were generated in the multi-photonic UCL bands, with a BFW of 22%, compared to <3% for the two-photonic 800nm emission (inset, Fig. 3e). These results support the hypothesis that the ability of UCNPs to generate new frequency components in response to modulated excitation is due to their inherent nonlinearity.

Investigations on other UCNPs, such as **YbEr@Y** nanoparticles, further validate the ability of UCNPs to function as modulation frequency mixers (Fig. S5a-f).

We see significant potential in exploiting the BF signal for practical applications, particularly compared to detection of the $2f$ signal. The BF signal can be generated at a sufficiently low arbitrary frequency, making it easily detectable also by low-speed detectors, e.g., cameras. However, in the UCNPs tested so far, the BFW of the usually dominant two-photonic UCL band is significantly smaller than its SHW. To address this issue, we turned to nanoparticle engineering to explore ways to increase the BFW.

We adopted the well-established core-(multi)shell strategy to regulate the photophysics of UCNPs.[17] Specifically, we expanded the optically active layers of $Tm^{3+}$-activated UCNPs with an additional $Yb^{3+}$-containing layer in the nanoparticle structure, resulting in NaYF$_4$:20%Yb$^{3+}$, 0.5%Tm$^{3+}$@NaYF$_4$: 20%Yb$^{3+}$@NaYF$_4$ (**YbTm@Yb@Y**) nanoparticles. Under identical experimental conditions, the 800nm emission of these UCNPs exhibited a significantly higher BF signal, compared to **YbTm@Y**, with a BFW reaching up to 14% (Fig. 3f). Noting that $Yb^{3+}$ ions in both the core and the shell regions absorb the excitation light and contribute to the upconversion process (mediated by energy migration across the extended $Yb^{3+}$ network and then by energy transfer to $Tm^{3+}$), these results suggest that this increased complexity of the upconversion pathways potentially enhances the nonlinearity of the UCNP emission and thereby the BFW.

To further test this hypothesis, we also synthesized $Er^{3+}$-activated UCNPs with different structural complexities, including NaYF$_4$:20%Yb$^{3+}$, 2%Er$^{3+}$@NaYF$_4$:20%Yb$^{3+}$ (**YbEr@Yb**) and NaYF$_4$:20%Yb$^{3+}$, 2%Er$^{3+}$@NaYF$_4$:20%Yb$^{3+}$@NaYF$_4$:30%Nd$^{3+}$, 20%Yb$^{3+}$ (**YbEr@Yb@YbNd**) nanoparticles, and investigated their responses to modulated excitation. As shown in Fig. S6, with an increased complexity in the nanoparticle structure and thereby in the photophysics, the ability to generate BF signals increased significantly in comparison to **YbEr@Y** UCNPs (from a BFW 6% for **YbEr@Y**, to 17% for YbEr@Yb, and to 20% for **YbEr@Yb@YbNd**, Fig. S6c). The BFW remained also relatively insensitive to changes in the average base modulation frequency (Fig. S6b). Moreover, by an additional excitation pathway, combining a 980nm (absorbed by $Yb^{3+}$) and an 808nm laser (absorbed by $Nd^{3+}$) modulated at different frequencies with a fixed difference of 2Hz, produced similarly robust and significant BF signals, with a likewise small dependence on the average base modulation frequency (Fig.

S6d). These results confirm that nanoparticle engineering, introducing an increased complexity in the upconversion pathways, offers opportunities to enhance modulation frequency mixing channels for use as BF signals.

Next, we explored the potential of the BF signal from UCNPs for background-free detection, particularly focusing on the detectability of the BF signal using low-speed cameras. We first conducted surface imaging experiments using a setup similar to that shown in Fig. S3 but with a sCMOS camera as the detector. A screw tap with a fluted '+' pattern, filled with **YbEr@Yb@YbNd** UCNPs, was used as the object. Two 980nm laser beams, modulated at 21 and 23Hz were overlapped for excitation. The sCMOS camera captured a time sequence of 2D images (frame exposure time: 10ms, 1000 frames in total) of the UCL signals (Fig. 4a).

As shown in Fig. 4b, the averaged image (over all acquired images in the sequence) to a large extent displays light collected from the surrounding region of the fluted '+' pattern, indicating substantial surface reflection of laser light despite the use of a 950nm short-pass filter (OD4) during detection. Next, we analyzed the data in the frequency domain. For each image in the sequence, the emission intensity was integrated across the entire field, yielding a summed intensity value. The generated time-trace of summed intensity values clearly reflects the excitation modulation (Fig. 4c), and FFT analysis revealed the BF signal at 2Hz, alongside signals at the two base modulation frequencies (Fig. 4d). For background suppression, we then frequency-filtered the 2Hz signal in the image sequence for each pixel and applied an inverse FFT (iFFT) to reconstruct the image (Fig. 4e). Comparing the intensity patterns in the averaged and frequency-filtered BF images reveals that the prominent residual excitation light present in the averaged image (Fig. 4b) is almost entirely suppressed in the BF image, while preserving the UCNP signal pattern (Fig. 4e). This effect is further highlighted in emission profiles across the fluted region (Fig. 4g). In contrast, filtering at either of the base modulation frequencies followed by image reconstruction does not achieve the same effect (Fig. 4f, Fig. S7). To further challenge the experimental conditions, we then introduced intense ambient light during the measurements. In the averaged image (Fig. 4h), the UCL signal pattern was then entirely obscured by the ambient light. However, the frequency-filtering approach at 2Hz, combined with iFFT, successfully recovered the UCL signal pattern (Fig. 4i), but not at 21Hz (Fig. 4j-k).

We then proceeded to tissue imaging studies. Four capillaries (1.5mm in diameter) filled with **YbTm@Yb@Y** suspensions at varying concentrations (0mg/mL, 2.5mg/mL, 5.0mg/mL, and 10.0mg/mL, respectively) were used as luminescent inclusions. Each capillary was embedded in chicken breast tissue at an approximate depth of 5 mm for imaging. The same dual beam modulated excitation (980nm) was applied to induce UCL. The excitation intensity was kept at approximately 460mW/cm², in compliance with the ANSI standard for laser safety in bioimaging applications.[18] The same image acquisition and data analysis procedures as above (Figs. 4a-g) were then used to obtain averaged and frequency-filtered BF images (Figs. 4l-q). As expected, due to light scattering through the 5mm thick tissue layer, the capillary structures were not sharply resolved in any of the recorded images (Fig. 4o-q). However, comparing intensity levels in the averaged and frequency-filtered BF images shows that residual excitation light is almost entirely suppressed in the BF image (Fig. 4p), whereas it contaminates the averaged image (Fig. 4o)—a result also highlighted in the image profiles (Fig. 4r). Filtering at either of the base modulation frequencies followed by image reconstruction does not produce the same effect (Fig. 4q-r, Fig. S8).

We further analyzed the signal strengths in both the averaged and 2Hz-filtered images for each capillary, focusing on the same regions of interest in the image/sample, as highlighted by green/blue boxes in Fig. 4s. A plot of the integrated signal strengths versus nanoparticle concentration in the capillaries shows that signals from the averaged images were significantly influenced by residual laser light, resulting in a substantial offset (Fig. 4t). Additionally, this laser-induced offset displayed marked spatial variations (Fig. 4t). In sharp contrast, the signals obtained from the 2Hz-filtered images were entirely free of residual laser light contamination, effectively recovering a linear relationship with nanoparticle concentration in each spot while preserving the actual relative spatial magnitudes (Fig. 4u).

**Conclusion**

UCNPs can function as frequency mixers in response to a modulated excitation intensity. Specifically, when exposed to sinusoidal/square-wave excitation, UCNPs can generate new frequency components that are not contained in the input excitation function. In the case of modulated excitation containing more than one base modulation frequency, they act as frequency mixers, producing difference frequency (or beating frequency-BF), as well as sum-frequency signals. The generation of the BF signal stems from the inherent nonlinear response of UCNPs to excitation intensity, and importantly, it can be significantly enhanced by increasing the complexity in the upconversion pathways via nanoparticle core-(multi)shell engineering. The BF signal of UCNPs offers a particularly attractive lock-in detection channel, detectable by low-speed cameras, with minimal background even in the presence of overwhelming background and (modulated) residual excitation light, as demonstrated by diffuse optical imaging experiments in this study.

Obtaining clean optical signals from the probes is crucial for bioimaging, especially for quantitative measurements.[19] Under challenging conditions, such as low probe concentration and deep tissue imaging, residual excitation light is often the primary source of contamination.[4] Our background-free detection approach has significant potential to overcome this issue and enable high-quality bioimaging.

**Associated Content**

**Data Availability Statement**

All relevant raw data behind this study are available via DOI: 10.5281/zenodo.14193471.

**Supporting Information**

The Supporting Information is available free of charge at XXX.

Section S1: Synthesis of nanoparticles; Section S2: Simulations of excitation modulation response of linear fluorophores; Section S3: Simulations of excitation modulation response of upconversion nanoparticles; Supplementary figures.

## Acknowledgements

This work was supported by the Swedish Foundation for Strategic Research (SSF, BENVAC RMX18-0041), the Swedish Research Council (VR 2021-04556), the Olle Engkvists Foundation (200-0514), the Carl Tryggers Foundation (CTS, 21: 1208, 23:2635), and the ÅForsk Foundation (23-322).

## References

1    Wei, Z. *et al.* Rare-earth based materials: an effective toolbox for brain imaging, therapy, monitoring and neuromodulation. *Light Sci Appl* **11**, 175, doi:10.1038/s41377-022-00864-y (2022).

2    Liu, K.-C. *et al.* A flexible and superhydrophobic upconversion-luminescence membrane as an ultrasensitive fluorescence sensor for single droplet detection. *Light Sci Appl.* **5**, e16136-e16136 (2016).

3    Xu, C. T. *et al.* Autofluorescence insensitive imaging using upconverting nanocrystals in scattering media. *Applied Physics Letters* **93**, doi:10.1063/1.3005588 (2008).

4    Zhu, B., Rasmussen, J. C., Lu, Y. & Sevick-Muraca, E. M. Reduction of excitation light leakage to improve near-infrared fluorescence imaging for tissue surface and deep tissue imaging. *Medical Physics* **37**, 5961-5970, doi:https://doi.org/10.1118/1.3497153 (2010).

5    Freudiger, C. W. *et al.* Label-Free Biomedical Imaging with High Sensitivity by Stimulated Raman Scattering Microscopy. *Science* **322**, 1857-1861 (2008).

6    Weibring, P., Edner, H. & Svanberg, S. Versatile Mobile Lidar System for Environmental Monitoring. *Appl. Opt.* **42**, 3583-3594 (2003).

7    Martinsons, C., Picard, N. & Carré, S. Optical Lock-in Spectrometry Reveals Useful Spectral Features of Temporal Light Modulation in Several Light Source Technologies. *LEUKOS* **19**, 146-164, doi:10.1080/15502724.2022.2077754 (2023).

8    Widengren, J., Schwille, Petra. Characterization of Photoinduced Isomerization and Back-Isomerization of the Cyanine Dye Cy5 by Fluorescence Correlation Spectroscopy. *J. Phys. Chem. A* **104**, 6416-6428 (2000).

9    Chmyrov, A., Sandén, T. & Widengren, J. Iodide as a Fluorescence Quencher and Promoter—Mechanisms and Possible Implications. *The Journal of Physical Chemistry B* **114**, 11282-11291, doi:10.1021/jp103837f (2010).

10   Spielmann, T., Xu, L., Gad, A. K. B., Johansson, S. & Widengren, J. Transient state microscopy probes patterns of altered oxygen consumption in cancer cells. *FEBS J.* **281**, 1317-1332 (2014).

11   Sanden, T., Persson, G., Thyberg, P., Blom, H. & Widengren, J. Monitoring Kinetics of Highly Environment Sensitive States of Fluorescent Molecules by Modulated Excitation and Time-Averaged Fluorescence Intensity Recording. *Anal. Chem.* **79**, 3330-3341, doi:10.1021/ac0622680 (2007).

12   Sanden, T., Persson, G. & Widengren, J. Transient State Imaging for Microenvironmental Monitoring by Laser Scanning Microscopy. *Anal. Chem.* **80**, 9589-9596, doi:10.1021/ac8018735 (2008).

13   Lakowicz, J. R., Laczko, G., Cherek, H., Gratton, E. & Limkeman, M. Analysis of fluorescence decay kinetics from variable-frequency phase shift and modulation data. *Biophys. J.* **46**, 463-477, doi:10.1016/S0006-3495(84)84043-6.

14   Pollnau, M., Gamelin, D. R., Lüthi, S. R., Güdel, H. U. & Hehlen, M. P. Power dependence of upconversion luminescence in lanthanide and transition-metal-ion systems. *Phys. Rev. B* **61**, 3337-3346 (2000).

15   Liu, H. *et al.* Balancing power density based quantum yield characterization of upconverting nanoparticles for arbitrary excitation intensities. *Nanoscale* **5**, 4770-4775 (2013).

16    Liu, H. *et al.* Deep tissue optical imaging of upconverting nanoparticles enabled by exploiting higher intrinsic quantum yield through using millisecond single pulse excitation with high peak power. *Nanoscale* **5**, 10034-10040 (2013).

17    Ji, Y. *et al.* Huge upconversion luminescence enhancement by a cascade optical field modulation strategy facilitating selective multispectral narrow-band near-infrared photodetection. *Light: Science & Applications* **9**, 184, doi:10.1038/s41377-020-00418-0 (2020).

18    Institute, A. N. S. *American national standard for safe use of lasers*. (Laser Institute of America, 2022).

19    Liu, H., Xu, C. T. & Andersson-Engels, S. Multibeam fluorescence diffuse optical tomography using upconverting nanoparticles. *Opt. Lett.* **35**, 718-720 (2010).

# Figures

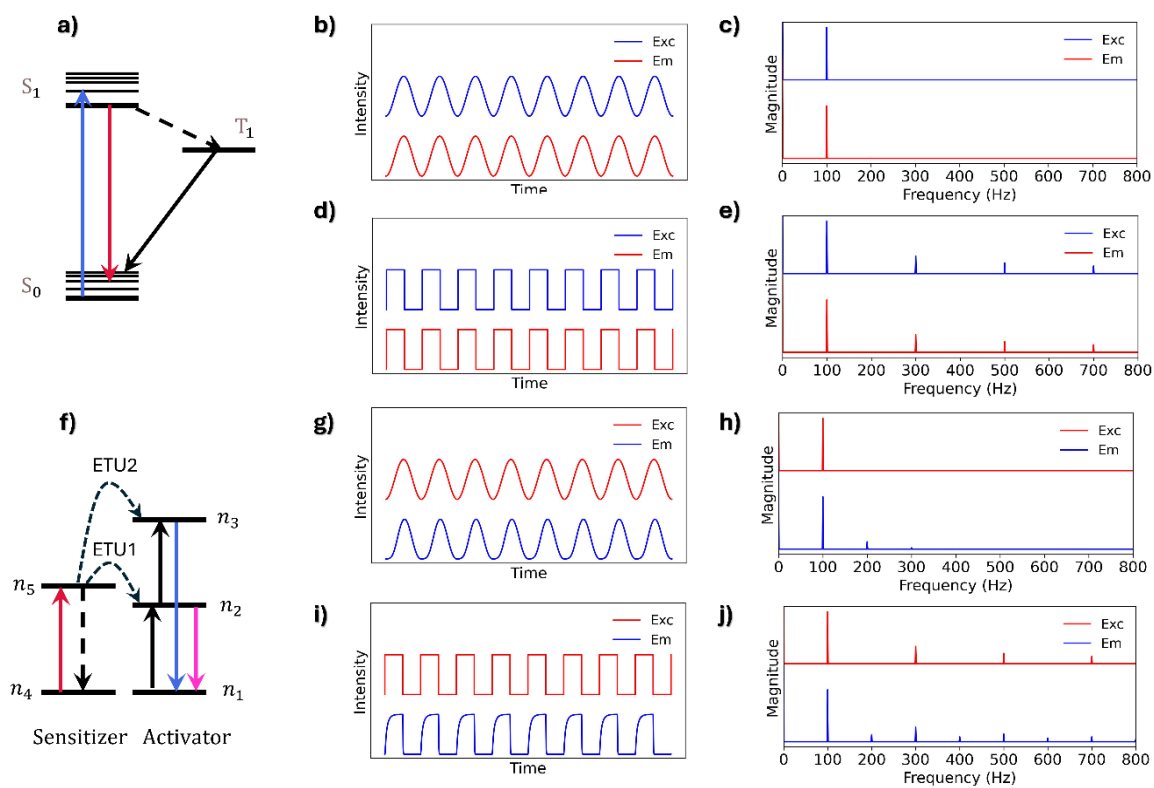

**Figure 1 Excitation modulation response of organic fluorophores and upconversion nanoparticles. (a)** Jablonski diagram of a typical organic fluorophore, showing light excitation, fluorescence, inter-system crossing (ISC), and triplet state relaxation processes (For rate equations and parameter values used in the simulations, see Section S2). Response of fluorophore fluorescence intensity to modulated excitations: **(b)** sinusoidal excitation, and **(d)** square-wave excitation with 50% duty cycle. The power spectra of the fast Fourier transform (FFT) of the fluorophore fluorescence signal compared to that of the excitation wave under **(c)** sinusoidal and **(e)** square-wave excitation with a 50% duty cycle. **(f)** Energy level diagram of a representative upconversion nanoparticle (UCNP), including a sensitizer and an activator yielding a two-step energy-transfer upconversion system (For rate equations and parameter values used in the simulations, see Section S3). Response of UCL intensity to modulated excitations: **(g)** sinusoidal excitation, and **(i)** square-wave excitation with a 50% duty cycle. The power spectra of the fast Fourier transform (FFT) of the UCL signal compared to that of the excitation wave under **(h)** sinusoidal and **(j)** square-wave excitation with 50% duty cycle.

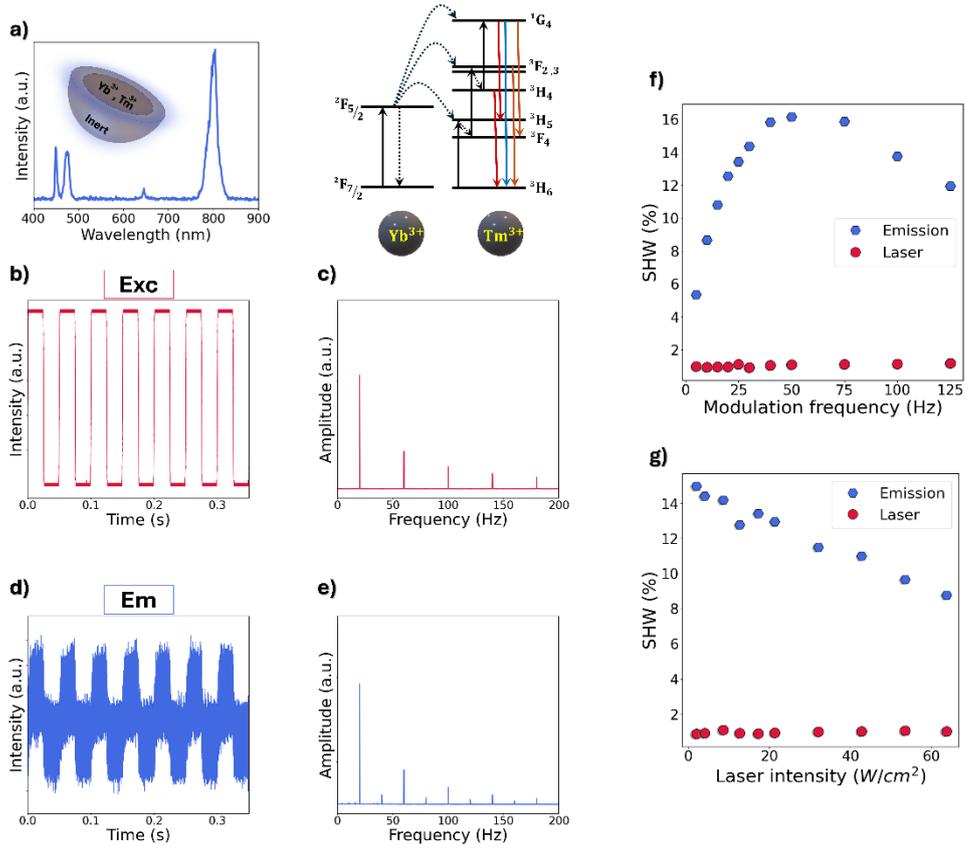

**Figure 2 Second-harmonic (SH) signal generation of upconversion nanoparticles (UCNPs) under square-wave excitation.** **(a)** UCL spectrum (Left) and photophysical model (Right) of the studied NaYF$_4$:20%Yb$^{3+}$, 0.5%Tm$^{3+}$@NaYF$_4$ (**YbTm@Y**) UCNPs under 980nm excitation. **(b)** The applied square-wave excitation (50% duty cycle with an excitation intensity within the pulses of 20.3W/cm$^2$) and **(c)** its power spectrum, obtained via a fast Fourier transform (FFT) from **(b)**. **(d)** The resulting UCL (430-850nm) and **(e)** its FFT power spectrum. The dependence of the defined SH weight (SHW) (eq. 1) of the UCL (430-850nm) on **(f)** the excitation modulation frequency and **(g)** the excitation intensity in the pulses.

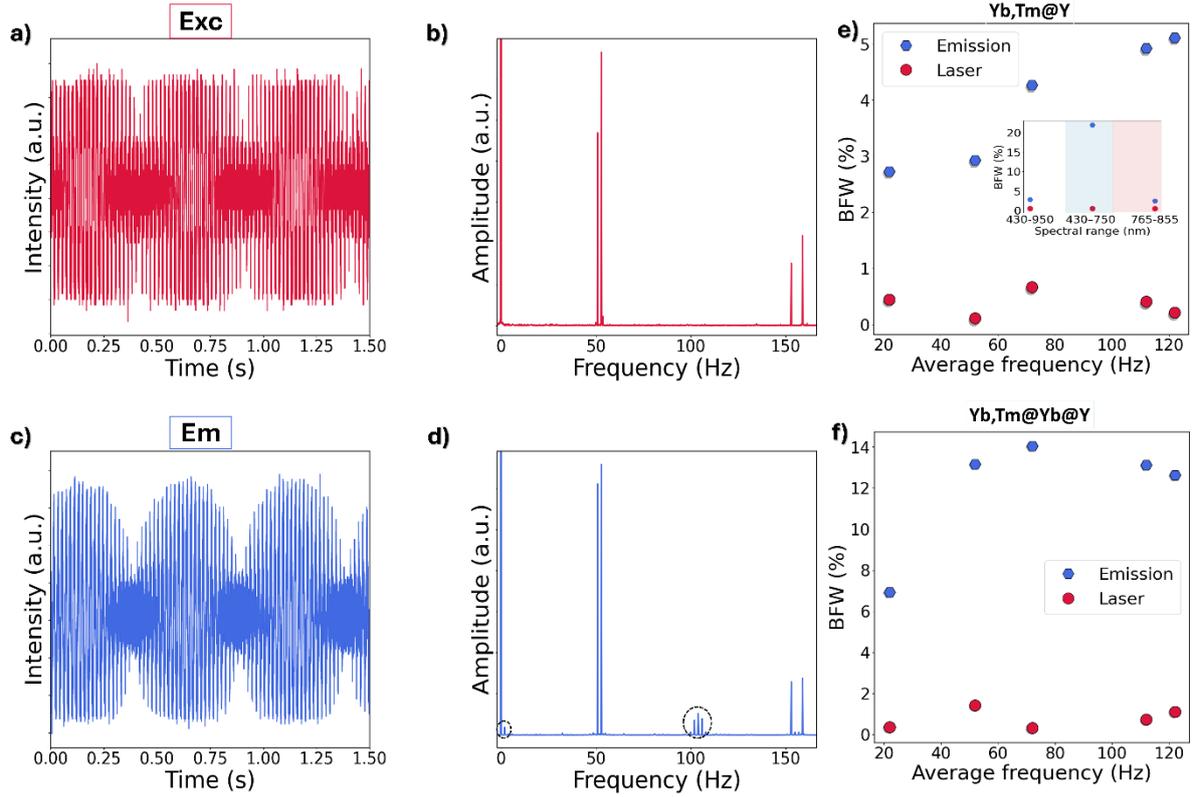

**Figure 3 Beating frequency (BF) signal generation of upconversion nanoparticles (UCNPs) under dual square-wave excitation. (a)** Recorded excitation intensity trace generated from the two merged 980nm laser beams modulated at 51 and 53Hz, respectively (excitation intensity: 30.5W/cm², each). **(b)** FFT power spectrum of the excitation intensity trace in **(a)**. **(c)** Intensity trace of the generated UCL signal (430-850nm) of **YbTm@Y** UCNPs, and **(d)** their FFT power spectrum. **(e)** The dependence of the defined BF weight (BFW) (eq. 2) of the UCL signal (430-850nm) of **YbTm@Y** UCNPs on the average modulation frequency (($f_1 + f_2)/2$). Inset: The BFWs of the UCL signal in different spectral ranges ($f_1 = 21$Hz and $f_2 = 23$Hz). **(f)** Similar plot as in **(e)**, now for the UCL signal (430-850nm) of NaYF$_4$:20%Yb$^{3+}$, 0.5%Tm$^{3+}$@NaYF$_4$: 20%Yb$^{3+}$@NaYF$_4$ (**YbTm@Yb@Y**) UCNPs.

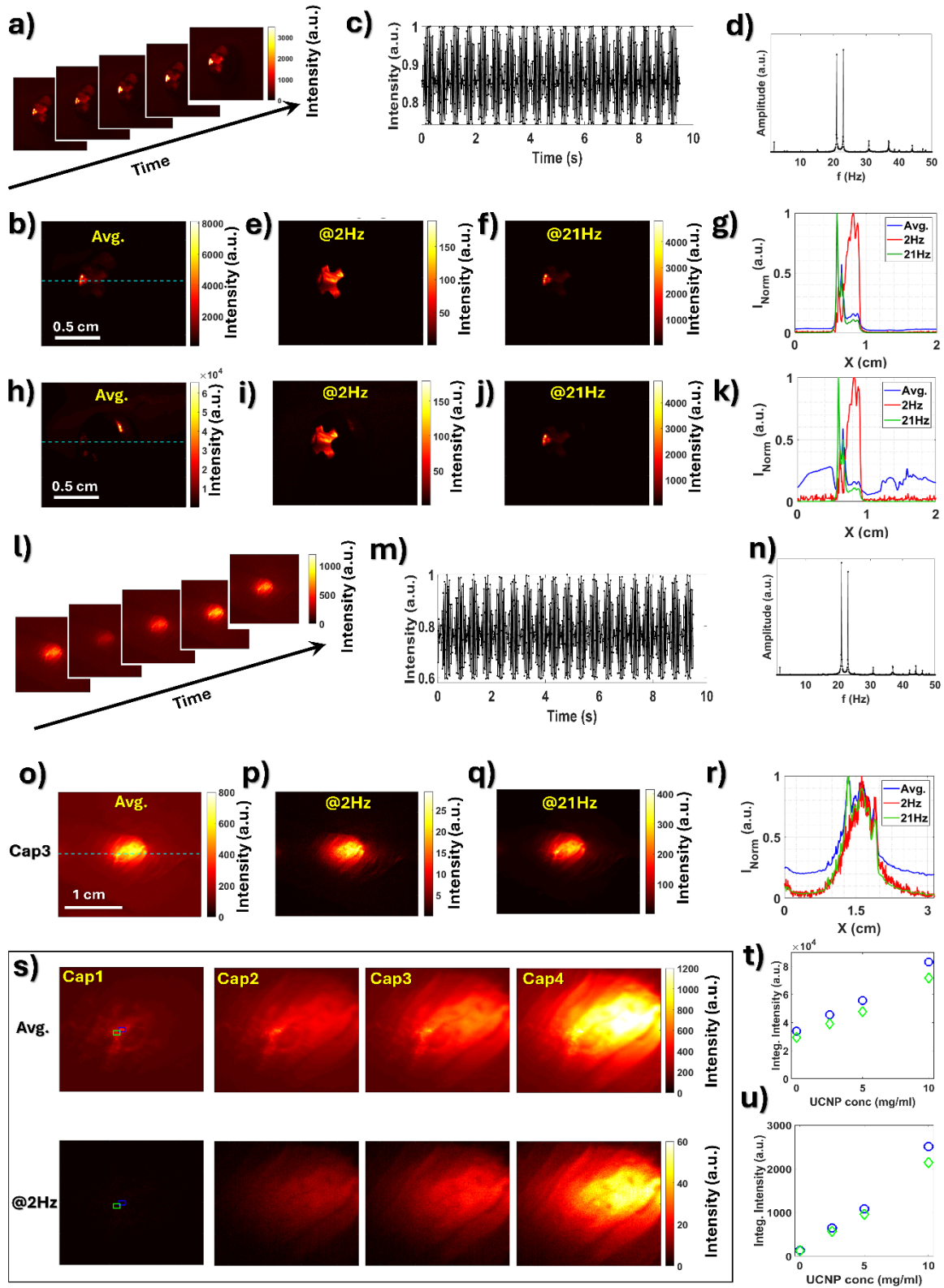

**Figure 4 Background-free wide-field imaging by detecting the BF signal of UCNPs. (a)** Image sequence of a **YbEr@Yb@YbNd** UCNP-capped screw tap under dual square-wave excitation of two 980nm laser beams (excitation intensity: 210mW/cm², each). **(b)** The averaged image of the UCNP-capped screw tap obtained from the recorded image sequence. **(c)** Time trace of the intensity, given by the integrated intensities in each of the images in the image sequence. **(d)** The FFT power spectrum of the time trace of integrated signals in **(b)**. **(e)** The BF (2Hz)-filtered image of the screw tap. **(f)** The 21Hz-filtered image of the screw tap. **(g)** Emission profiles along the marked line in the averaged, 21Hz-filtered and 2Hz-filtered images. **(h)** The averaged image of the UCNP-capped screw tap of the image sequence recorded with intense ambient light. **(i)** The BF (2Hz)-filtered image of the screw tap of the image sequence recorded with intense ambient light. **(j)** The 21Hz-filtered image of the screw tap of the image sequence recorded with intense ambient light. Intensities between images **(b)**, **(e)**, **(f)**, **(h)**, **(i)**, **and (j)** can be compared. **(k)** Emission profiles along the marked line in the averaged, 21Hz-filtered and 2Hz-filtered images. **(l)** Image sequence of a capillary filled with **YbTm@Yb@Y** (capillary 3, 5mg/mL) embedded in chicken breast tissue at an approximate depth of 5mm under dual square-wave excitation of two 980nm laser beams (excitation intensity: 230mW/cm$^2$, each). **(m)** Time trace of integrated intensity values from the image sequence in **(l)**. **(n)** The FFT power spectrum of the time trace of integrated signals in **(m)**. **(o)** The averaged image of the capillary obtained from the recorded image sequence in **(l)**. **(p)** The BF (2Hz)-filtered image of capillary 3. **(q)** The 21Hz-filtered image of capillary 3. **(r)** Emission profiles along the marked line in the averaged, 21Hz-filtered and 2Hz-filtered images. Profiles in **(g)**, **(k)**, and **(r)** are normalized to their maximum values. **(s)** The averaged and 2Hz-filtered images for capillaries filled with different concentrations of the same UCNPs (capillary 1: 0mg/mL, capillary 2: 2.5mg/mL, capillary 3: 5mg/mL, capillary 4: 10mg/mL. Intensities between images **(o)**, **(p)**, **(q)**, and **(s)** can be compared. **(t), (u)** Signals from the averaged and 2Hz-filtered images in two selected regions (marked with green or blue boxes in **(s)**) as a function of the UCNP concentration within the capillaries.

**Supporting Information**

# Lanthanide upconversion nonlinearity: a key probe feature for background-free deep-tissue imaging

Niusha Bagheri[a], Chenyi Wang[b], Du Guo[a], Anbharasi Lakshmanan[a], Qi Zhu[a], Nahid Ghazyani[c], Qiuqiang Zhan[b], Georgios A. Sotiriou[d], Haichun Liu*[a], Jerker Widengren*[a]

*[a] Experimental Biomolecular Physics, Department of Applied Physics, KTH Royal Institute of Technology, SE-106 91, Stockholm, Sweden*

*[b] Centre for Optical and Electromagnetic Research, South China Academy of Advanced Optoelectronics, South China Normal University, Guangzhou 510006, P. R. China*

*[c] Faculty of Physics, Kharazmi University, Tehran, Iran.*

*[d] Department of Microbiology Tumor and Cell Biology Karolinska Institute, SE-171 77, Stockholm, Sweden*

*Corresponding authors: haichun@kth.se; jwideng@kth.se

**Contents:**

**Section S1. Synthesis of nanoparticles**

**Chemicals and Materials**

Anhydrous yttrium(III), ytterbium(III), neodymium(III), erbium(III), and thulium(III) chlorides (99.9%), oleic acid (90%), octadec-1-ene (90%), methanol, sodium hydroxide, ammonium fluoride were purchased from Sigma-Aldrich (St. Louis, MO, USA). Cyclohexane was purchased from VWR (Radnor, PA, USA).

**Synthesis of core NaYF$_4$:Yb$^{3+}$,Er$^{3+}$ and core NaYF$_4$:Yb$^{3+}$,Tm$^{3+}$ nanoparticles**

The core nanoparticles were synthesized according to previously reported protocols.[1] Into a 100-mL three-neck round-bottom flask, 1 mmol of lanthanide chlorides, oleic acid (6 mL), and octadec-1-ene (15 mL) were mixed. To prepare NaYF$_4$:Yb$^{3+}$,Er$^{3+}$, 0.78 mmol YCl$_3$, 0.2 mmol YbCl$_3$, and 0.02 mmol ErCl$_3$ were used. To prepare NaYF$_4$:Yb$^{3+}$,Tm$^{3+}$, 0.795 mmol YCl$_3$, 0.2 mmol YbCl$_3$ and 0.01 mmol TmCl$_3$ were used. In each case, the mixture was heated at 160°C for 30 min with stirring under continuous Ar flow until the mixture became homogeneous, and then it was cooled down to room temperature. After cooling, a methanolic solution of NaOH (2.5 mmol) and NH$_4$F (4 mmol) was added to the reaction mixture. The reaction temperature gradually increased from room temperature to 300°C. Different reaction time was used to prepare different nanoparticles. Next, the mixture was naturally cooled down to room temperature, and the NaYF$_4$:Yb$^{3+}$,Er$^{3+}$ or NaYF$_4$:Yb$^{3+}$,Tm$^{3+}$ nanoparticles were collected by centrifugation. After centrifugation, the solution above a white pellet (precipitate of nanoparticles) was discarded and the nanoparticles were dispersed in cyclohexane (10 mL), precipitated by ethanol (5 mL), and centrifugated. This step was repeated one more time and, finally, the nanoparticles were dispersed in cyclohexane for subsequent use.

**Synthesis of core-shell NaYF$_4$:Yb$^{3+}$,Er$^{3+}$@NaYF$_4$:x%Yb$^{3+}$ and NaYF$_4$:Yb$^{3+}$,Tm$^{3+}$@NaYF$_4$:x%Yb$^{3+}$ nanoparticles**

Core-shell NaYF$_4$:Yb$^{3+}$,Er$^{3+}$@NaYF$_4$:x%Yb$^{3+}$ and NaYF$_4$:Yb$^{3+}$,Tm$^{3+}$@NaYF$_4$:x%Yb$^{3+}$ nanoparticles were synthesized according to previously reported protocols with modifications.[1] Stoichiometric lanthanide chloride shell precursors, oleic acid (6 mL), and octadec-1-ene (15 mL) were mixed and heated at 160°C for 30 min under Ar flow to get a homogeneous yellowish solution and cooled down. After cooling, the suspension of pre-synthesized core nanoparticles, a methanolic solution of NaOH (1.25 mmol) and NH$_4$F (2 mmol) were added to the reaction mixture and heated gradually to 300°C for 1 h. The mixture was cooled down to room temperature and the core-shell nanoparticles were collected by centrifugation. After centrifugation, the top solution was discarded and the core-shell nanoparticles left were dispersed in cyclohexane (10 mL), precipitated by ethanol (5 mL), and centrifugated. This step was repeated one more time and, finally, the core-shell nanoparticles were dispersed in cyclohexane for subsequent use.

**Synthesis of core-shell-shell NaYF$_4$:20%Yb$^{3+}$, 2%Er$^{3+}$@NaYF$_4$:20%Yb$^{3+}$@NaYF$_4$:30%Nd$^{3+}$, 20%Yb$^{3+}$ nanoparticles**

NaYF$_4$:20%Yb$^{3+}$, 2%Er$^{3+}$@NaYF$_4$:20%Yb$^{3+}$ nanoparticles were first synthesized and then used as the seed nanoparticles to grow a NaYF$_4$:30%Nd$^{3+}$, 20%Yb$^{3+}$ layer, following the procedure described above.

## Section S2. Simulations of excitation modulation response of linear fluorophores

The electronic state population dynamics of a linear fluorophore, as depicted in the Jablonski diagram in Fig. 1a, are modeled by the following rate equations:

$$\frac{\mathrm{d}[S_0]}{\mathrm{d}t} = -k_{01}P(t)[S_0] + k_{10}[S_1] + k_T[T_1] \tag{S1}$$

$$\frac{\mathrm{d}[S_1]}{\mathrm{d}t} = k_{01}P(t)[S_0] - k_{10}[S_1] - k_{isc}[S_1] \tag{S2}$$

$$\frac{\mathrm{d}[T_1]}{\mathrm{d}t} = k_{isc}[S_1] - k_T[T_1] \tag{S3}$$

Here, $[S_0]$, $[S_1]$, and $[T_1]$ denote the population probabilities of the singlet ground state ($S_0$), the singlet excited state ($S_1$), and the triplet state ($T_1$), respectively; $k_{01}$ is the peak excitation rate, $k_{10}$ is the radiative decay rate of the $S_1$ state, $k_{isc}$ is the inter-system crossing rate, and $k_T$ is the decay rate of the $T_1$ state; $P(t)$ represents the peak-normalized excitation function, which can take different forms, harmonic wave or square wave, in simulating the response of the fluorophore to different optical stimuli. The parameter values used in the simulations of Figs. 1c and 1e are summarized in Table S1. The equations (S1-S3) were solved using an open-source software COPASI, using the initial conditions $[S_0](t=0) = 1$, and $[S_1](t=0) = [T_1](t=0) = 0$.

**Table S1.** Parameter values used in the simulations of excitation modulation response of linear fluorophores

| $k_{01}$ (s$^{-1}$) | $k_{10}$ (s$^{-1}$) | $k_{isc}$ (s$^{-1}$) | $k_T$ (s$^{-1}$) |
|---|---|---|---|
| $1.2 \times 10^7$ | $2.5 \times 10^8$ | $1 \times 10^7$ | $5 \times 10^5$ |

**Section S3. Simulations of excitation modulation response of upconversion nanoparticles**

Numerical simulations of the response of upconversion nanoparticles to modulated excitation were performed based on a simplified two-photon upconversion model depicted in Fig. 1f.

The kinetics of this upconversion system are modeled by the following rate equations:

$$\frac{dn_1}{dt} = -W_{51}n_1n_5 + k_{21}n_2 + k_{31}n_3 \tag{S4}$$

$$\frac{dn_2}{dt} = W_{51}n_1n_5 - W_{52}n_2n_5 - k_{21}n_2 \tag{S5}$$

$$\frac{dn_3}{dt} = W_{52}n_2n_5 - k_{31}n_3 \tag{S6}$$

$$\frac{dn_4}{dt} = -k_{45}P(t)n_4 + k_{54}n_5 + W_{51}n_1n_5 + W_{52}n_2n_5 \tag{S7}$$

$$\frac{dn_5}{dt} = k_{45}P(t)n_4 - k_{54}n_5 - W_{51}n_1n_5 - W_{52}n_2n_5 \tag{S8}$$

where $n_i$ is the population density of energy state $i$ ($i$ = 1-5) in Fig. 1f; $k_{45}$ is the excitation pumping rate for the transition of state **4** → state **5**; $k_{54}$, $k_{21}$ and $k_{31}$ are the decay rates for state **5**, state **2**, and state **3**, respectively; $W_{51}$ and $W_{52}$ are the coefficients for ETU1 and ETU2 (see Fig. 1f), respectively; $P(t)$ represents the peak-normalized excitation function. The parameter values used for the simulations of UCNPs are summarized in Table S2.

**Table S2.** Parameter values used in the simulations of excitation modulation response of upconversion nanoparticles

| $k_{45}$ (s$^{-1}$) | $k_{54}$ (s$^{-1}$) | $k_{21}$ (s$^{-1}$) | $k_{31}$ (s$^{-1}$) | $W_{51}$(cm$^3$ s$^{-1}$) | $W_{52}$(cm$^3$ s$^{-1}$) |
|---|---|---|---|---|---|
| 7.9 | $1 \times 10^3$ | $7.6 \times 10^2$ | $1.1 \times 10^4$ | $2 \times 10^{-16}$ | $2 \times 10^{-15}$ |

The equations (S4-S8) were solved using an open-source software COPASI, using the initial conditions $[n_1](t=0) = 1.25 \times 10^{20}$ cm$^{-3}$, $[n_4](t=0) = 1.52 \times 10^{21}$ cm$^{-3}$, and $[n_2](t=0) = [n_3](t=0) = [n_5](t=0) = 0$ cm$^{-3}$.

## Supplementary figures

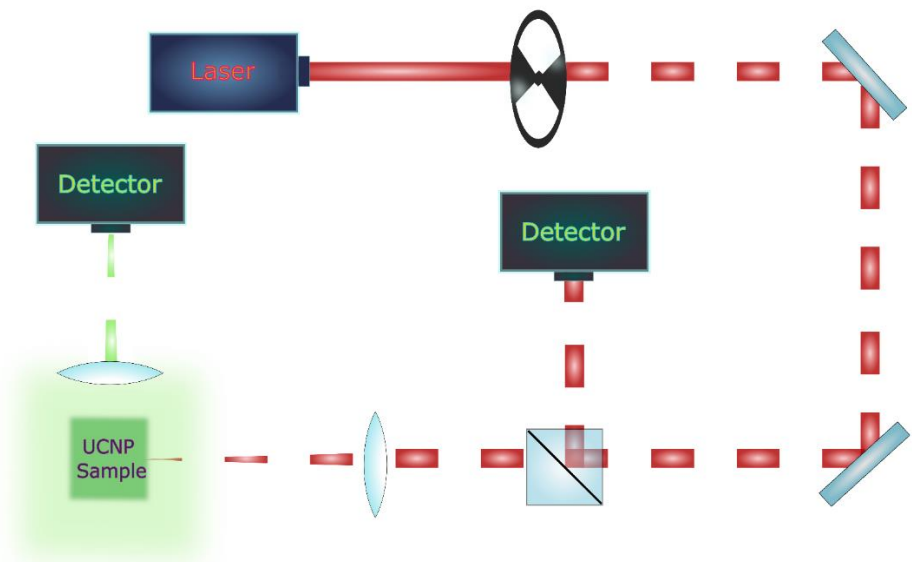

**Figure S1 Optical setup to investigate the response of UCNPs to a single-beam square-wave excitation with a 50% duty cycle.** Herein, a continuous-wave (CW) laser at 980 nm was used as the excitation source, and a mechanical chopper modulated the excitation light to generate the desired square wave. The laser beam was split into two paths by a cube beamsplitter: one path was monitored by a photodiode to track the excitation wave, while the other path illuminated the UCNP sample, generating upconversion emissions that were detected by an avalanche photodiode (APD). The chopper allowed us to vary the modulation frequency.

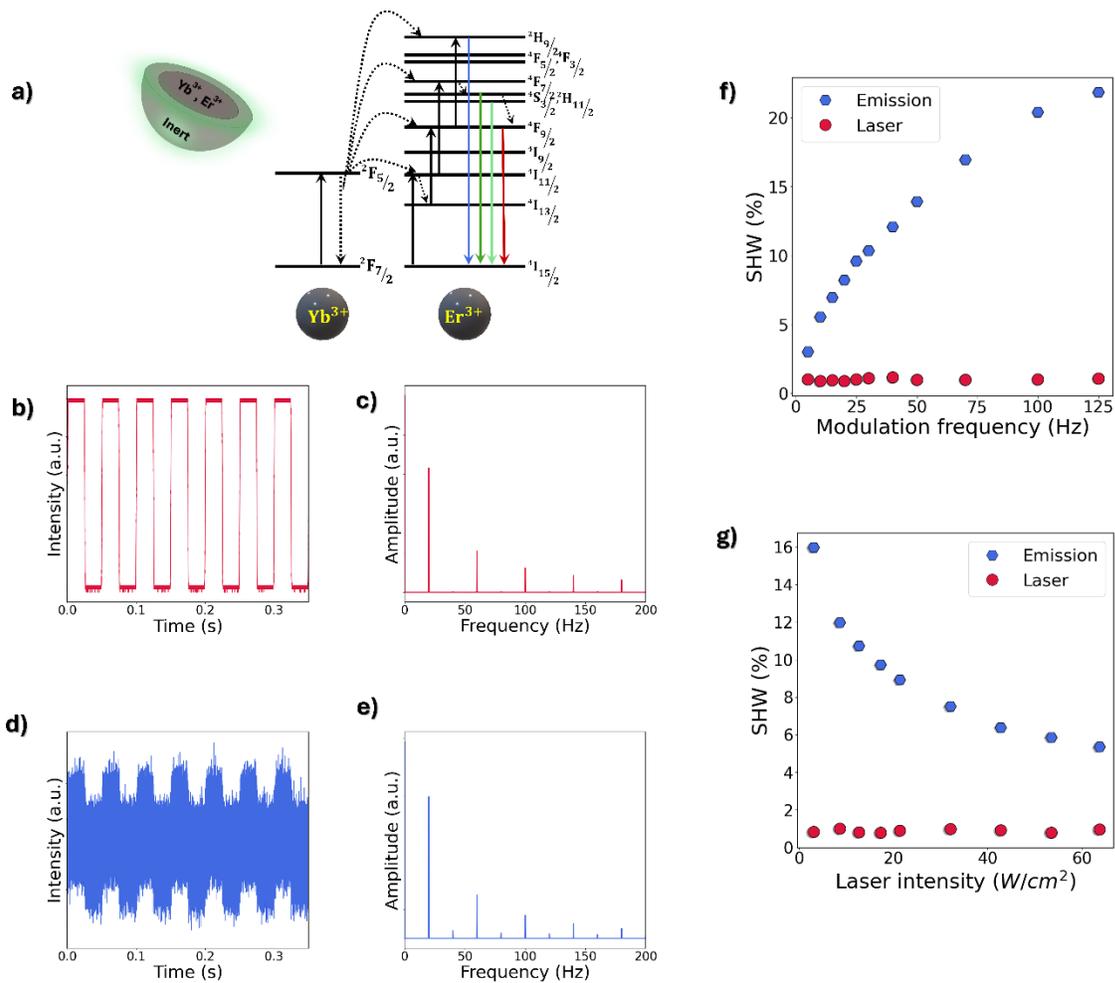

**Figure 2 Second-harmonic (SH) signal generation of NaYF₄:20%Yb³⁺, 2%Er³⁺@NaYF₄ (YbEr@Y) UCNPs under square-wave excitation.** **(a)** Upconversion nanoparticle structure (Left) and population mechanism (Right) of the studied **YbEr@Y** UCNPs **(b)** The applied square-wave excitation with a 50% duty cycle (maximum excitation intensity: 20.3 W/cm²) and **(c)** its power spectrum of the fast Fourier transform (FFT). **(d)** The generated upconversion emission (500-700 nm) and **(e)** its FFT power spectrum. The dependence of the SH weight (SHW) of the upconversion emission (500-700 nm) on **(f)** excitation modulation frequency and **(g)** excitation intensity.

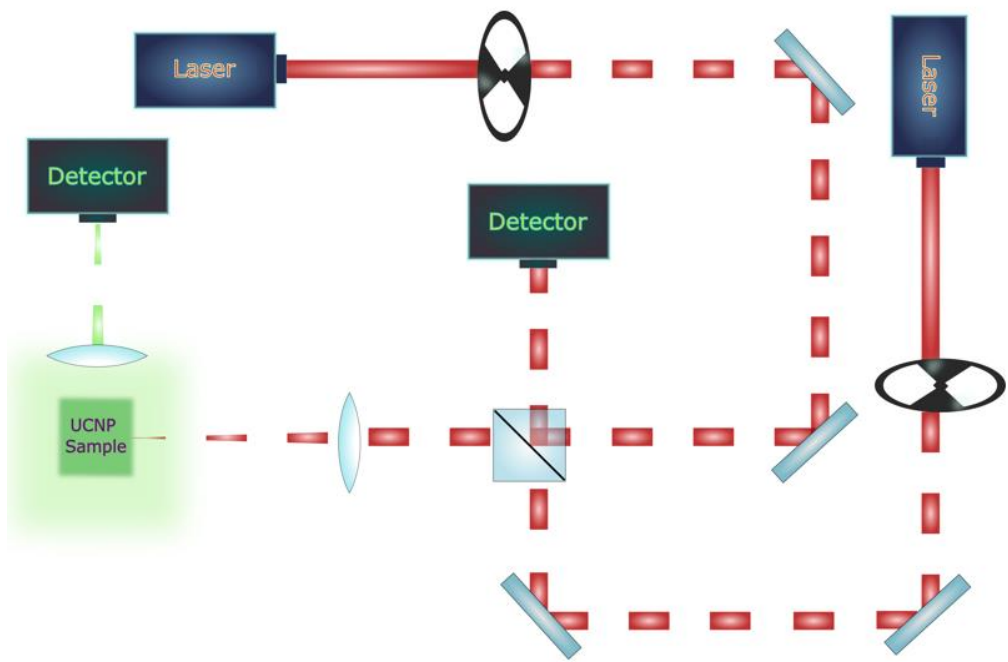

**Figure S3 Optical setup to investigate the response of UCNPs to dual beam square-wave excitation both with a 50% duty cycle.** Herein, two continuous-wave (CW) lasers at selected wavelengths (808 or 980 nm) were used as the excitation sources, each modulated by a mechanical chopper to generate the desired excitation square wave. The combined excitation wave was monitored by a photodiode, while the generated upconversion emissions were detected by an avalanche photodiode (APD). The choppers allowed us to vary the modulation frequencies.

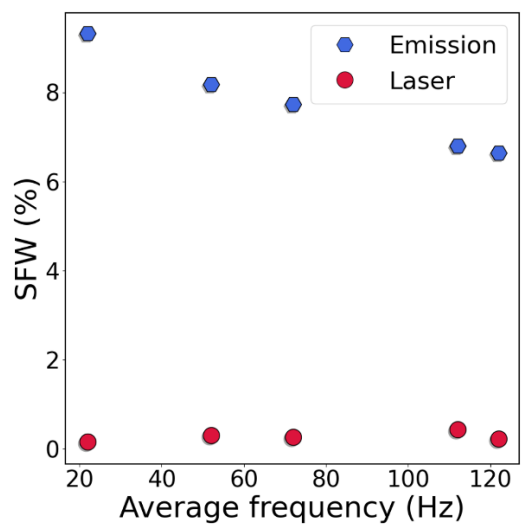

**Figure S4** The dependence of the sum-frequency signal weight of the upconversion emission signal (430-850 nm) of NaYF$_4$:20%Yb$^{3+}$, 0.5%Tm$^{3+}$@NaYF$_4$ (**YbTm@Y**) UCNPs on the average modulation frequency ($(f_1 + f_2)/2$) under dual square-wave excitation

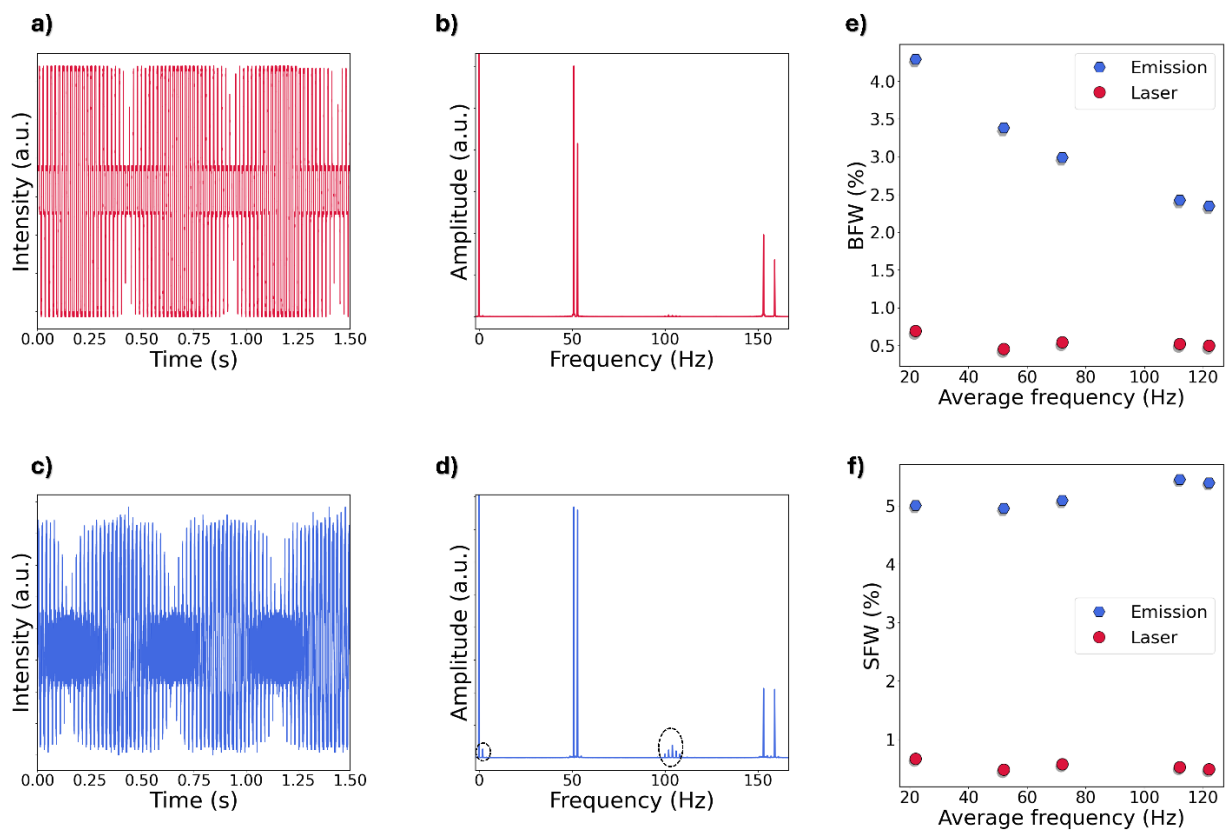

**Figure S5 Beating frequency (BF) signal generation of NaYF$_4$:20%Yb$^{3+}$, 2%Er$^{3+}$@NaYF$_4$ (YbEr@Y) UCNPs under dual square-wave excitation. (a)** The recorded excitation light signal from two merged 980 nm laser beams modulated at 51 and 53 Hz, respectively (excitation intensity: 30.5 W/cm$^2$, each), and **(b)** its FFT power spectrum. **(c)** The recorded generated upconversion emission signal (500-700 nm) of **YbEr@Y** UCNPs, and **(d)** its FFT power spectrum. The dependence of **(e)** the defined BF weight (BFW) (eq. 2) and **(f)** the sum-frequency signal weight of the upconversion emission signal (500-700 nm) of **YbEr@Y** UCNPs on the average modulation frequency (($f_1 + f_2$)/2).

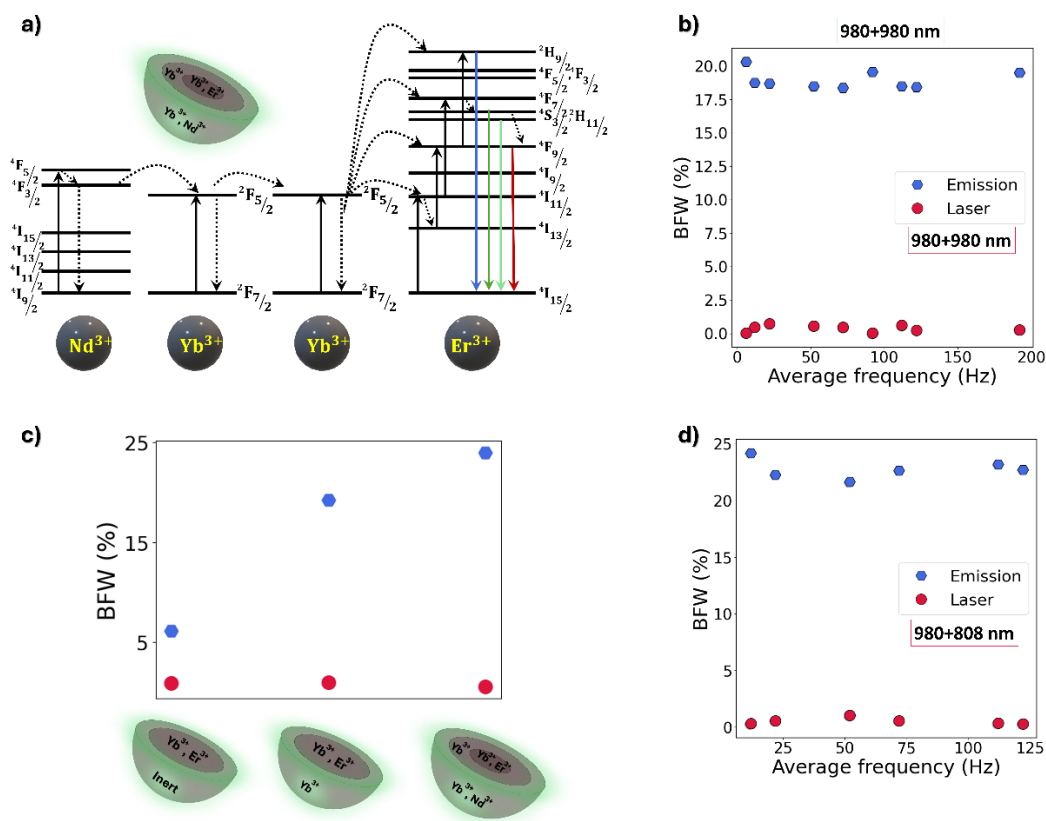

**Figure S6 Optimization of the BF signal of UCNPs through nanoparticle engineering. (a)** Model describing upconversion emission pathways of NaYF$_4$:20%Yb$^{3+}$, 2%Er$^{3+}$@NaYF$_4$:20%Yb$^{3+}$@NaYF$_4$:30%Nd$^{3+}$, 20%Yb$^{3+}$ (**YbEr@Yb@YbNd**) core-multishell UCNPs under 980 or 808 nm excitation. **(b)** Dependence of the BFW of the upconversion emission signal (500-700 nm) of **YbEr@Yb@YbNd** UCNPs on the average modulation frequency (($f_1 + f_2$)/2) under dual square-wave excitation of two 980 nm laser beams (50% duty cycle, $f_1 - f_2$ kept at 2 Hz, excitation intensity within pulses: 30.5 W/cm$^2$, each). **(c)** Comparison of the BFWs of the **YbEr@Y**, **YbEr@Yb**, and **YbEr@Yb@YbNd** UCNPs under dual square-wave excitation of two 980 nm laser beams (excitation intensity within pulses: 30.5 W/cm$^2$, each). **(d)** Dependence of the BFW of the upconversion emission signal (500-700 nm) of the same UCNPs and under the same conditions as in **(b)**, but now with one of the laser beams emitting at 980 nm replaced by a laser beam emitting at 808 nm.

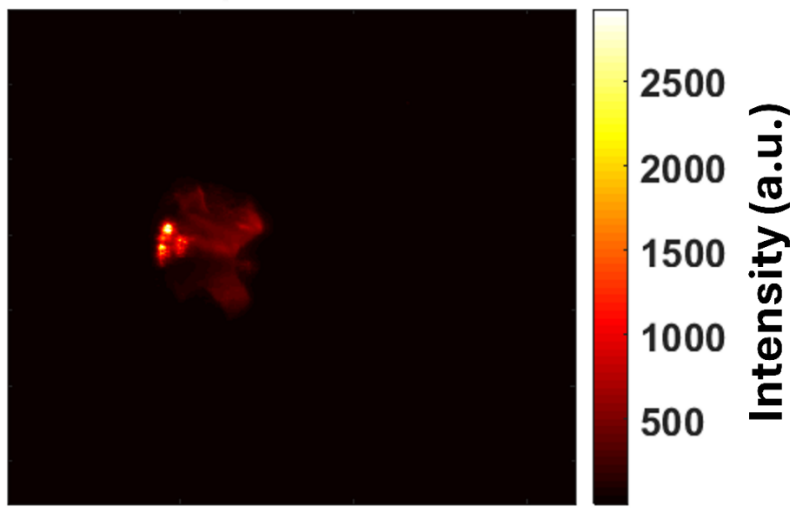

**Figure S7** The 23 Hz-filtered image of the screw tap capped by NaYF$_4$:20%Yb$^{3+}$, 2%Er$^{3+}$@NaYF$_4$:20%Yb$^{3+}$@NaYF$_4$:30%Nd$^{3+}$, 20%Yb$^{3+}$ (**YbEr@Yb@YbNd**) UCNPs.

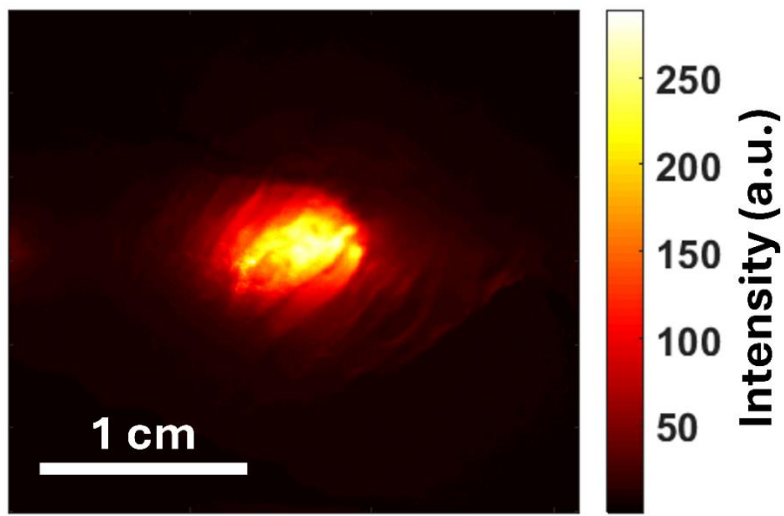

**Figure S8** The 23 Hz-filtered image of capillary 3, filled with NaYF$_4$:20%Yb$^{3+}$, 0.5%Tm$^{3+}$@NaYF$_4$: 20%Yb$^{3+}$ (**YbTm@Yb**) UCNPs (5 mg/mL)